\documentclass[pra,reprint,showpacs,superscriptaddress]{revtex4-1}

\usepackage{graphicx}
\usepackage{bm}
\usepackage{color}
\usepackage{float}
\usepackage[USenglish]{babel}
\usepackage{amsmath}
\usepackage{siunitx}
\usepackage{multirow}
\usepackage{braket}
\usepackage{hyperref}
\hypersetup{
	colorlinks=true,
	linkcolor=blue,
	citecolor=blue,
	urlcolor=blue
}
\begin{document}
	
	
	\title{Observation of dipolar splittings in high-resolution atom-loss spectroscopy of $^6$Li $p$-wave Feshbach resonances}
	
	\author{Manuel Gerken}
	\author{Binh Tran}
	\author{Stephan H\"afner}
	\affiliation{Physikalisches Institut, Universit\"at Heidelberg, Im Neuenheimer Feld 226, 69120 Heidelberg, Germany}
	
	\author{Eberhard Tiemann}
	\affiliation{Institut f\"ur Quantenoptik, Leibniz Universit\"at Hannover, Welfengarten 1, 30167 Hannover, Germany}
	
	\author{Bing Zhu}
	\email{bzhu@physi.uni-heidelberg.de}
	\author{Matthias Weidem\"uller}
	\email{weidemueller@uni-heidelberg.de}
	\affiliation{Physikalisches Institut, Universit\"at Heidelberg, Im Neuenheimer Feld 226, 69120 Heidelberg, Germany}
	\affiliation{Hefei National Laboratory for Physical Sciences at the Microscale and Department of Modern Physics, and CAS Center for Excellence and Synergetic Innovation Center in Quantum Information and Quantum Physics, University of Science and Technology of China, Hefei, Anhui 230026, China}

	\date{\today}

	\begin{abstract}
		
		We report on the observation of dipolar splitting in $^6$Li $p$-wave Feshbach resonances by high-resolution atom-loss spectroscopy. The Feshbach resonances at 159 G and 215 G exhibit a doublet structure of \SI{10}{mG} and \SI{13}{mG}, respectively, associated with different projections of the orbital angular momentum. The observed splittings agree very well with coupled-channel calculations. We map out the temperature dependence of the atom-loss spectrum allowing us to extrapolate resonance positions and the corresponding widths to zero temperature. The observed dipolar splitting in fermionic lithium might be useful for the realization of the quantum phase transition between the polar and axial $p$-wave superfluid phases.
		
	\end{abstract}
		
	\maketitle
	
	\section{Introduction}
	
	It is generally known that magnetic $p$-wave Feshbach resonances (FRs) in ultracold atomic collisions exhibit a dipolar splitting, resulting in two resonance features corresponding to $m_l=0$ and $m_l=\pm1$  \cite{Ticknor2004, Strauss2010}. Here, $m_l$ is the projection of the orbital angular momentum $l$ along the external magnetic field. This non-degeneracy was predicted to give rise to a phase transition from a $p_x$ to a $p_x+ip_y$ state in a $p$-wave superfluid (SF) \cite{Gurarie2007}, which plays an essential role in SF liquid $^3$He \cite{Volovik1992} and neutron SFs inside neutron stars \cite{Page2011, Shternin2011}. The dipolar splitting was observed in a spin-polarized gas of $^{40}$K \cite{Ticknor2004} and a large variety of quantum-gas mixtures \cite{Pilch2009a, WangP2011, Repp2013, WangF2013, Dong2016}. It is attributed to effective spin-spin (\emph{ss}) interactions including the magnetic dipole-dipole interaction and the second-order spin-orbit coupling \cite{Strauss2010}. Recently, the observation of dipolar splitting has been extended to $d$-wave resonances \cite{Cui2017, Yao2019}, and an additional splitting mechanism has been discovered for $m_l=+1$ and $m_l=-1$ due to spin-rotation interaction in $p$-wave FRs \cite{Zhu2019, Haefner2019}.
	
	Due to the favorable properties of its FRs, the fermionic isotope of lithium, $^6$Li, one of the only two stable fermionic isotopes among the alkali metals, has been a major workhorse in experiments with quantum-degenerate Fermi gases.  Prime applications include the study of the BCS-BEC crossover \cite{Jochim2003, Zwierlein2003,Bartenstein2004, Chin2004, Bourdel2004, Zwierlein2004, Bartenstein2004a}, the Efimov effect \cite{Ottenstein2008, Huckans2009, Wenz2009, Williams2009, Lompe2010}, the double SF \cite{Ferrier-Barbut2014}, and the Fermi-Hubbard model in optical lattices \cite{Chin2006, Parsons2015, Duarte2015, Omran2015, Greif2016, Mitra2018}. Although dipolar splitting in $^6$Li $p$-wave FRs was predicted to be $\sim$ \SI{10}{mG} \cite{Chevy2005}, it was not resolved so far due to insufficient magnetic resolution in previous experiments \cite{Zhang2004, Schunck2005, Nakasuji2013, Yoshida2017, Waseem2017, Waseem2018}. By using an optical lattice $p$-wave Feshbach molecules in the $m_l=\pm1$ component were selectively formed without resolving the dipolar splitting \cite{Waseem2016}, while the studies of the molecular binding energy and lifetime in Refs. \cite{Fuchs2008, Inada2008, Maier2010} neglected this effect.
	
	In this work we report high-resolution trap-loss spectroscopy on three $^6$Li $p$-wave FRs in the two lowest-energy hyperfine states. Dipolar splittings of about \SI{10}{mG} are resolved in two of the three resonances and systematic effects of the sample temperature on the observed splitting are investigated. We model the observations with a coupled-channels (cc) calculation and find an excellent agreement  between experiment and modeling.
	

	\section{Dipolar splitting}
	\label{sec:expt}
	
	\begin{figure*}[t]
		\centering
		\includegraphics[width=\textwidth]{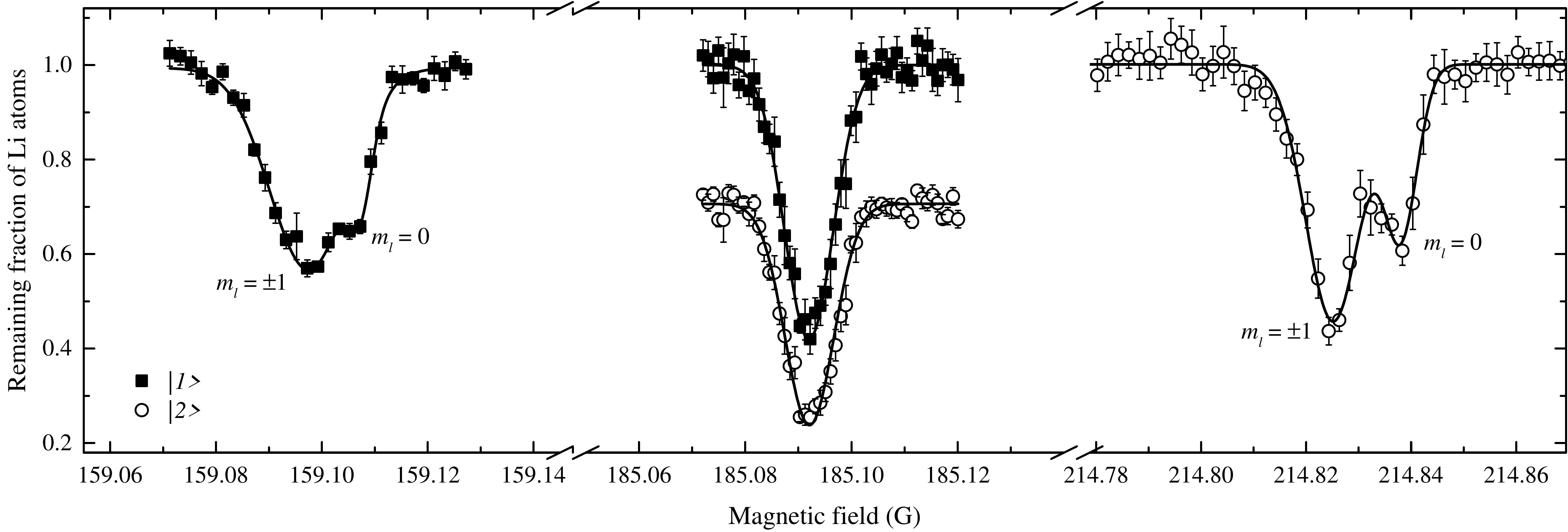}
		\caption{$^6$Li $p$-wave FRs in the $\ket{1}\oplus\ket{1}$, $\ket{1}\oplus\ket{2}$, and $\ket{2}\oplus\ket{2}$ entrance channels observed as atomic loss features. The figure shows the fractional remaining Li atoms in spin-state $\ket{1}$ (solid squares) or $\ket{2}$ (open circles) after a holding time of \SI{500}{ms}, \SI{150}{ms} and \SI{100}{ms} from left to right, respectively. The resonances near \SI{159}{G} in the $\ket{1}\oplus\ket{1}$ channel and \SI{214}{G} in the $\ket{2}\oplus\ket{2}$ channel show doublet structures, where quantum numbers $m_l$ are assigned according to theoretical modeling. The data of spin state $\ket{2}$ for the resonance near \SI{185}{G} is vertically shifted to avoid overlapping with that of $\ket{1}$. The solid curves are fits of multi-peak Gaussian functions from which the resonance positions $B^e$ and widths $w^e$ are extracted as listed in Table \ref{tab:pwaves}.}
		\label{fig:pwaves}
	\end{figure*}

	A detailed description of our experimental apparatus has already been given in previous publications \cite{Repp2013,Pires2014}. In brief, we optically trap a laser-cooled sample of $^6$Li atoms in the two energetically lowest hyperfine states $\ket{f=1/2,m_f=1/2}$ (labeled as $\ket{1}$) and $\ket{f=1/2,m_f=-1/2}$ (labeled as $\ket{2}$) in a cigar-shaped optical dipole trap. Here, $f$ and $m_f$ refer to the total angular momentum and its projection along the external magnetic field. The sample is evaporatively cooled within \SI{\sim5}{s} at a magnetic field of $\sim$ \SI{890}{G}. Both spin states can be selectively populated by removing the other one with a short resonant light pulse. Finally, we end up with $3\times 10^4$ atoms in each spin state at a temperature of $T\approx$ \SI{0.14}{\micro K}. The trapping frequencies are determined to be $\omega=2 \pi \times$ \SI{(30, 170, 180)}{Hz}, resulting in a peak atomic density of $1.6(1)\times$\SI{10^{11}}{cm^{-3}} and $T/T_F=0.53(4)$.
	
	Three $^6$Li $p$-wave FRs are identified by performing magnetic-field dependent loss spectroscopy of Li samples either polarized at a single hyperfine state ($\ket{1}$ or $\ket{2}$) or in a mixture of both spin states, as shown in Fig. \ref{fig:pwaves}. After optimized holding times of \SI{500}{ms} (\SI{159}{G} FR), \SI{150}{ms} (\SI{185}{G} FR) and \SI{100}{ms} (\SI{214}{G} FR) the remaining number of Li atoms in each spin-state is detected by absorption imaging. The points for each resonance are taken in a random order of the magnetic field. Calibration of the magnetic field is performed  by rf-spectroscopy of the nuclear spin-flip transition between the Li $\ket{1}$ and $\ket{2}$ state. The Breit-Rabi formula is then used to infer the magnetic field strength. Its total uncertainty is derived from calibration measurements to be \SI{10}{mG}, resulting from daily drifts, residual field curvature along the long axis of the atom cloud, and calibration uncertainties. 
	
	In Fig. \ref{fig:pwaves}, doublet structures corresponding to $m_l=\pm 1$ and $m_l=0$ components in the $^6$Li $p$-wave FRs near \SI{159}{G} and \SI{214}{G} are presented. Gaussian functions are fitted to the data to extract the experimental loss peak positions $B^e_{m_l}$ and loss widths $w^e_{m_l}$. Doublet splittings $\delta^e$ are defined as distances between the two peaks. The results are listed together with the theoretical resonance splitting $\delta^t$ obtained from a full cc calculation in Table \ref{tab:pwaves}.
	
	The observed doublet splittings are in excellent agreement with a cc calculation including an effective spin-spin interaction, similar to that in Ref. \cite{Knoop2009}. We measure a splitting of \SI{10(1)}{mG} and \SI{13(1)}{mG} for the \SI{159}{G} and \SI{214}{G} $p$-wave FR, respectively. In case of the $p$-wave FR in channel $\ket{1}\oplus\ket{2}$, the small calculated splitting of \SI{4}{mG} is not resolved in the experiment, since the splitting is considerably smaller than the typical width of the loss peak. The assignments of quantum numbers $m_l$ in Fig. \ref{fig:pwaves} are from the cc calculation. Our observations also agree very well with previous predictions \cite{Chevy2005, Fuchs2008}.

	\begin{table}[t]
		\caption{$^6$Li $p$-wave FRs. The experimentally obtained resonance positions $B^e_{\pm1}$, doublet splittings $\delta^e$ and widths $w^e_{m_l}$ are extracted by fitting multi-peak Gaussian functions to the loss spectra. The first number in brackets give their determination uncertainty and the second number the systematic uncertainty of \SI{10}{mG}. The theoretical doublet splittings $\delta^t$ obtained from the cc scattering calculations for a relative kinetic energy of $k_B\times$ \SI{140}{nK}, matching with the experimentally measured temperature, are given.}
		\label{tab:pwaves}
		\begin{ruledtabular}
			\begin{tabular}{l|ccccc}
				& $B^e_{\pm1}$ (G) & $w^e_0$ (mG) & $w^e_{\pm1}$ (mG) & $\delta^e$ (mG) & $\delta^t$ (mG) \\\hline
				$\ket{1}\oplus\ket{1}$	&159.097(1)(10) & 6(2) & 17(2) & 10(1) & 10\\
				$\ket{1}\oplus\ket{2}$	&185.092(1)(10) & 10(1) & 10(1) & - & 4 \\
				$\ket{2}\oplus\ket{2}$	&214.825(1)(10) & 7(2) & 10(1) & 13(1) & 13\\			
			\end{tabular}
		\end{ruledtabular}	
	\end{table}
	
	\section{Effects of finite temperature}
	\label{sec:effects}
	
	\begin{figure*}[t]
		\centering
		\includegraphics[width=\textwidth]{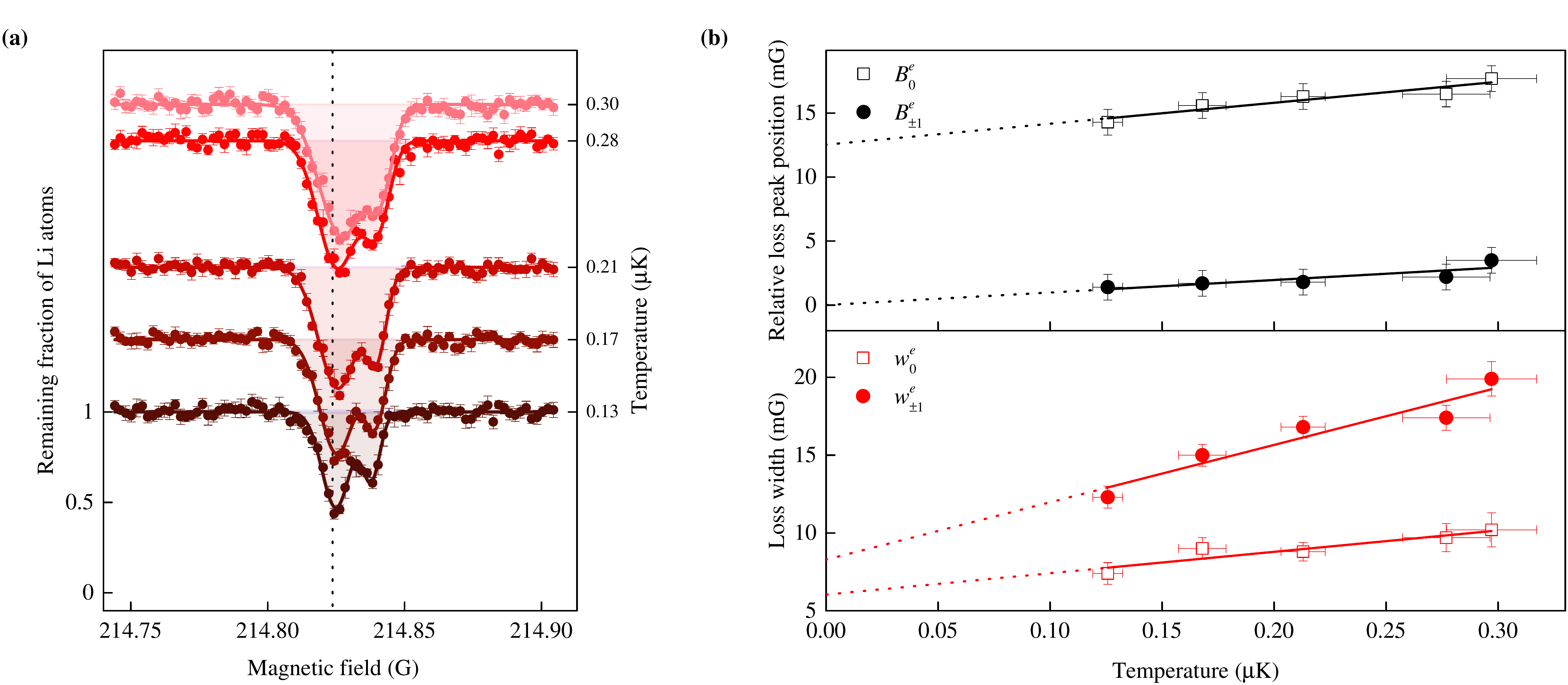}
		\caption{Temperature dependence of trap loss spectrum near the \SI{214}{G} $^6$Li $p$-wave FR. (a) Loss spectrum at different sample temperatures. The base-lines of the profiles are shifted according to $T$, as shown at the right axis. The data is fitted using a double-Gaussian function to extract the loss peak positions $B_{m_l}^e$ and widths $w^e_{m_l}$. The dotted vertical line shows the extrapolated peak position of $B^e_{\pm1}$ at zero temperature at \SI{214.824}{G}. (b) $B^e_{m_l}$ (black symbols) and $w^e_{m_l}$ (red symbols) of the observed loss features at different $T$ for the $m_l=\pm 1$ (solid circles) and $m_l=0$ (open squares) components. $B^e_{m_l}$ are referenced to the zero temperautre value of $B^e_{\pm1}$ at \SI{214.824}{G}. The lines are linear fits to the data.} 
		\label{fig:pwaves_temp}
	\end{figure*}
	
	The finite temperature of the gas causes systematic effects on Feshbach resonances in atom-loss spectra, such as asymmetries, shifts, broadenings, and saturation \cite{Ticknor2004,Dong2016, Zhang2004, Waseem2017, Waseem2018}. Averaging over the Maxwell-Boltzmann distribution of the collision energy results in an asymmetric lineshape \cite{Ticknor2004, Chevy2005} shifting the resonance center and broadening the spectrum. Saturation of atom loss stems from the unitarity limit at finite temperatures, rendering it impossible to resolve magnetic resonance splittings below magnetic fields of order $3k_BT/2\delta\mu$ \cite{Waseem2017, Waseem2018}, with  $\delta\mu$ denoting the relative magnetic moment between the molecular and atomic states. To ensure that our measurements (magnetic field resolution of $\sim$ \SI{1}{mG}) are not limited by such effects, temperatures below $T\sim$ \SI{0.7}{\mu K} near the \SI{214}{G} resonance ($\delta\mu=k_B\times$ \SI{118(8)}{\mu K/G} \cite{Fuchs2008}) are required.
	
	We investigate the temperature effects on the resonance lineshapes near the \SI{214}{G} $p$-wave FR.
	Evaporation is stopped at varying trap depths resulting in $T\le$ \SI{0.3}{\mu K}, and atomic loss spectra are recorded, as shown in Fig. \hyperref[fig:pwaves_temp]{2(a)}. Above \SI{0.3}{\mu K} the line-shape becomes asymmetric and the doublet spitting is unresolvable due to the increase of both the temperature-broadened width and the influence of effects due to unitarity, as discussed above. The extracted loss peak positions and widths are plotted in Fig. \hyperref[fig:pwaves_temp]{2(b)}. 
	
	As expected \cite{Ticknor2004}, the resonance peak moves towards lower magnetic field at a lower temperature. We notice that the observed loss width $w^e$ is considerably larger than the expected temperature-broadened width $k_BT/\delta\mu$, ranging between \SI{1.1}{mG} and \SI{2.5}{mG} within the temperature range of the experiments. To the lowest order, the observed temperature dependence of the loss peak position and width can be well approximated by linear functions (black and red lines in Fig. \hyperref[fig:pwaves_temp]{2(b)}). This yields for the resonance positions slopes of \SI{16(3)}{mG/\mu K} and \SI{10(3)}{mG/\mu K} for the $m_l=0$ and $m_l=\pm1$ components, respectively. Although the slopes are slightly different for the two $|m_l|$ values, the doublet splitting $\delta^e$ remains almost constant within the experimental uncertainty when $T$ is varied. Extrapolation to zero temperature gives a splitting of \SI{13(1)}{mG}. 
	
	Using the Breit-Wigner theoretical approach described in Refs. \cite{Waseem2017, Waseem2018} and assuming $k_BT$ as the smallest energy scale, we have estimated the shift of the resonances for two- and three-body loss processes to be $5k_B/2\delta\mu=$ \SI{21.2}{mG/\mu K} and $k_B/\delta\mu=$ \SI{8.5}{mG/\mu K}, respectively. By comparing the experimental and theoretical slopes we infer that both two- and three-body loss contributes to the observed temperature dependence. Disentangling these loss channels by time-resolved measurements is beyond the scope of this paper. In order to realize a $p$-wave SF, one would have to mitigate loss by identifying appropriate temperature and density regimes, or by using dimensional control \cite{Zhou2017, Waseem2017}.
	
	For the observed widths of the loss resonances, we obtain slopes of \SI{37(7)}{mG/\mu K} and \SI{14(4)}{mG/\mu K} for the $m_l=\pm1$ and $m_l=0$ components, while the zero-temperature widths are \SI{8(1)}{mG} and \SI{6(1)}{mG}, respectively. Contributions to the observed widths include the resonance intrinsic width $\gamma$ \cite{Chevy2005, Waseem2017, Waseem2018}, the thermal broadening, and other experimental broadening effects like the trap-induced density inhomogeneity and magnetic-field noise. However, from the observed linearity and the fact that $w_e(T)>k_BT/\delta\mu$ in Fig. \hyperref[fig:pwaves_temp]{2(b)} one may infer that $\gamma\gg k_BT$ in our experiment. This conclusion, however, is in disagreement with recent estimations of $\gamma$ extracted from the temperature- and interaction-strength-dependent two- \cite{Waseem2017} and three-body \cite{Waseem2018} loss rate constants near the $p$-wave resonances above the Fermi temperature. 
	
	%
	%
	%

	\section{Conclusion}
	\label{sec:conclusion}
	
	In conclusion, we have performed high-resolution atom-loss spectroscopy of $p$-wave Feshbach resonances in an optically trapped ultracold $^6$Li gas and resolved the splittings of about \SI{10}{mG} between $m_l=0$ and $m_l=\pm1$ components. Our measurements agree excellently with a full cc calculation including spin-spin interactions. 
	In spin-polarized Fermi gases near $p$-wave FRs, there have been predictions of the phase transition from a polar $p_x$ state to an axial $p_x+ip_y$ state as well as the topological transition from a gapless to a gapped $p_x+ip_y$ phase \cite{Botelho2005,Gurarie2005,Cheng2005}. In the past, stabilities of single-component Fermi gases and $p$-wave Feshbach molecules near $p$-wave FRs were investigated, while the realization of $p$-wave SF in quantum gases is still challenging due to severe inelastic losses \cite{Regal2003a, Zhang2004, Schunck2005, Gaebler2007, Fuchs2008, Inada2008, Maier2010}. Dimensional confining has been identified as a promising route for stabilizing Fermi gases near $p$-wave FRs and realizing $p$-wave SF pairing \cite{Zhou2017, Waseem2017}. Resolving the predicted dipolar splitting of FRs in $^6$Li provides a good starting point for future investigations in this direction exploring the individual control of $m_l=0,\pm1$ scattering processes \cite{Guenter2005}. Furthermore, experimental studies of the $p$-wave contact can now be extended to $^6$Li, which was, previously, only accessible in a $^{40}$K gas due to its larger splitting of the $m_l$ components \cite{Luciuk2016}.  

	\begin{acknowledgements}
		We are grateful to J. Ulmanis, S. Jochim and S. Kokkelmans for inspiring discussions. S.H. acknowledges support by the IMPRS-QD. This work is supported in part by the Heidelberg Center for Quantum Dynamics, the DFG/FWF Research Unit FOR2247 under Project No. WE2661/11-1, and the DFG Collaborative Research Centre SFB1225 (ISOQUANT).
	\end{acknowledgements}

	\bibliography{Mixtures}     

\end{document}